\documentclass{aa}

\usepackage{graphicx}
\usepackage[varg]{txfonts}
\usepackage{amssymb,amsmath}
\usepackage{eurosym}
\usepackage{hyperref}
\hypersetup{
    colorlinks=true,
    linkcolor=blue,
    filecolor=magenta,      
    urlcolor=cyan,
    pdftitle={Sharelatex Example},
    bookmarks=true,
    pdfpagemode=FullScreen,
    }
    
\usepackage{natbib}
\bibpunct{(}{)}{;}{a}{}{,} 
\usepackage{mathabx}

\newcommand{\mbf}[1]{\mathbf{#1}}
\newcommand{\mrm}[1]{\mathrm{#1}}

\begin{document}

   \title{Matrix-propagator approach to compute fluid Love numbers and applicability to extrasolar planets}

   \author{S. Padovan\inst{1}\fnmsep\thanks{ \email{sebastiano.padovan@dlr.de}},
          T. Spohn \inst{1},
          P. Baumeister \inst{1,2},
          N. Tosi \inst{1,2},
          D. Breuer \inst{1},
          Sz. Csizmadia \inst{1},
          H. Hellard \inst{1},
          \and
          F. Sohl \inst{1}
          }

   \institute{Institute of Planetary Research, German Aerospace Center (DLR), Berlin, Germany
         \and
         Department of Astrophysics, Technische Universit{\"a}t, Berlin, Germany}

  \abstract
   {The mass and radius of a planet directly provide its 
   bulk density, which can be interpreted in terms of its 
   overall composition.
   Any measure of the radial mass distribution provides 
   a first step in
   constraining the interior structure.
   The fluid Love number $k_2$ provides such a measure, 
   and estimates of $k_2$ for extrasolar planets are expected 
   to be available in the coming years thanks to improved 
   observational facilities 
   and the ever-extending temporal baseline of 
   extrasolar planet observations. 
}
   {We derive a method for calculating the Love numbers $k_n$ of any 
   object given its density profile, which is routinely 
   calculated from 
   interior structure codes.}
   {We used the matrix-propagator technique, a method frequently 
   used in the geophysical community.}
   {We detail the calculation and apply it to the case of 
   \object{GJ 436b}, 
   a classical example of the degeneracy of mass-radius relationships,    to illustrate how measurements of $k_2$ can improve 
   our understanding of the 
   interior structure of extrasolar planets. 
   We implemented the method in a code that  
   is fast, 
   \href{https://bitbucket.org/sebastianopadovan/planetary-fluid-love-numbers/src/master/}{freely available}, 
   and easy to combine with preexisting 
   interior structure codes.
    While the linear approach presented here 
    for the calculation of the Love numbers 
    cannot treat the presence of 
    nonlinear effects that may arise under certain 
    dynamical conditions, 
    it is applicable to close-in gaseous 
    extrasolar planets like hot Jupiters, 
    likely the first targets for which $k_2$ will be measured.}
   {}

   \keywords{planets and satellites: interiors --
   planets and satellites: fundamental parameters -- 
   planets and satellites: gaseous planets --
   planets and satellites: terrestrial planets
   planets and satellites: individual: GJ 436b
               }
\titlerunning{Matrix-propagator approach to compute fluid Love numbers}
\authorrunning{S. Padovan et al.}
   \maketitle
%

\section{Introduction}\label{Sec__Intro}
A knowledge of the mass and radius of an exoplanet allows determining its
mean density, which is the most basic indicator of its composition.
Using an example from the solar system, 
the similarity of the densities of the Earth and Mercury 
along with the diversity of their sizes allows us to infer, under the assumption that
they are both composed of rocks and metals, that the metallic component in Mercury
is larger than in the Earth \citep{Ash1971}.
Planets are roughly spherical objects because self-gravitation overcomes
material strength for bodies larger than about a few hundred kilometers.
However, a spherical shape does not guarantee a differentiated interior.
An integral measure of the concentration of mass---and thus, indirectly of 
differentiation and interior structure---is provided by the 
normalized moment of inertia (MoI), 
defined for a spherical body of mass $M$ and radius $R$  as
\begin{equation}
{\rm MoI}=\frac{1}{MR^2}\int_0^V\rho(r)r^2_{\perp}dV,
\label{Eq__MoI}
\end{equation}
where $V$ is the volume, $\rho(r)$ is the density as a function of the radius $r$, and 
$r_{\perp}$ is the distance from an axis passing through the center
of mass of the body.
Planets are not perfectly spherical, and the value of the MoI
will in general depend on the chosen axis.
However, there are only three independent moments of inertia,
usually indicated with $A,$ $B,$ and $C$, and 
the MoI as defined in equation (\ref{Eq__MoI}) can be 
taken as representing the normalized mean moment of 
inertia given by $(A+B+C)/(3MR^2)$. 
A homogeneous spherical body has a value of MoI of $0.4$, while
in a gravitationally stable body where density increases with depth,
$0 \leq {\rm MoI} <0.4$, with 0 representing the value for a point mass.

The most common causes of departure 
from sphericity at large spatial scales for planets are
rotation and gravitational interactions with other bodies: parent star, 
other planets, and moons.
In general it is possible to express these perturbations in terms of a 
potential. 
In the case of a tide-inducing perturber of mass $M_{\rm p}$ at a distance $d,$ 
the tide-generating potential $W$ within the planet
can be expanded in spherical harmonics as
\citep[e.g.,][]{Kaula1966}
\begin{eqnarray}
W\left(r,\psi\right)=\frac{GM_{\rm p}}{d}\sum_{n=2}^{\infty}\left(\frac{r}{d}\right)^{n}P_{n}\left(\cos\psi\right)=\sum_{n=2}^{\infty}W_{n},
\label{Eq__TideInducingPotential}
\end{eqnarray}
where $r$ is the coordinate of a point within the planet, 
$\psi$ is the angle with respect to the center of mass
between the point and the perturber, 
and $P_{n}$ is the Legendre polynomial of degree $n$.
Similarly, a rotational potential can be written as a harmonic
term of degree 2 \citep[e.g.,][]{Murray1999}.
Under the assumption that the response of the planet to these
perturbing potentials is linear, each harmonic degree of the 
perturbing potential $W$ will generate a response with the 
same degree, $V_{n}^{\rm p}$. At the surface,
\begin{equation}
V^{\rm p}_{n}=k_{n}W_{n},
\label{Eq__TidalPotential}
\end{equation}
where $k_{n}$ is the gravitational
Love number, introduced by A. E. H. Love
\citep[although not with this name,][]{Love11},
and can be regarded as a useful way
of condensing in a single parameter the many unknowns
controlling the gravitational response of the planet to the 
perturbation.
In addition to $k_{n}$, the Love number
$h_{n}$
describes the radial
displacement of the surface that results from the 
presence of the perturbing potential.
The equipotential surface is defined by the
external potential $W_{n}$ and the additional
potential $k_{n}W_{n}$,
corresponding to $(1+k_{n})W_{n}/g_0,$
with $g_0$ the gravitational acceleration at the surface.
For a fluid planet the location of the surface, 
which is an equipotential surface,
corresponds to the radial displacement
$h_{n}W_{n}$, thus resulting in 
the simple relation $h_{n}=1+k_{n}$
\citep{Munk1960}.

In general, the forward calculation of the Love numbers 
for a given planetary interior structure requires 
the knowledge of the density, viscosity, and rigidity
profiles within the planet, in addition to the timescale of the perturbing
potential \citep[e.g.,][]{Alterman1959}.
However, if the planet is in hydrostatic equilibrium, that is, if it
responds as a fluid, only the 
density profile is required \citep[e.g.,][]{Sterne1939,Gavrilov1975}.
Thus, both the moment of inertia and the fluid Love numbers 
only depend on the distribution of matter in the interior of the planet,
and there exists an equation, known as the Darwin-Radau equation,
which provides a link between the MoI and $k_2$ (or $h_2$).
However, this relation is only an approximation 
\citep{Murray1999,Kramm2011,Helled2011}, and in this
paper, the MoI and the Love numbers are calculated 
independently, using equation (\ref{Eq__MoI})
and the method presented 
in Section \ref{Sec__MatrixPropagator},
respectively.

The expressions we derive for the Love numbers $k_{n}$ 
are based on the matrix-propagator method, which seeks the solution
to a system of differential equations through the use of matrices, and
traces back to the ideas of \citet{Thomson1950} and \citet{Haskell1953},
which in turn are part of the theoretical framework developed 
by \citet{Volterra1887}, as illustrated by \citet{Gilbert1966}.
Here, we apply a procedure similar to that developed by \citet{Wolf1994}.
The matrix-propagator method requires 
the knowledge of the radial density profile, discretized at a given
number of internal interfaces, without requiring the knowledge
(or numerical calculation) 
of the local derivative of the density profile.

The paper is structured as follows. 
In section \ref{Sec__PotentialMethod}
we illustrate the calculation of the Love numbers $k_{n}$ for the simple case
of a homogeneous planet or a planet with two constant density layers,
which we then generalize to the
case of a planet defined by any number of layers
in section \ref{Sec__MatrixPropagator}.
As an illustration of the method, in section \ref{Sec__GJ436b}
we apply the theory to the case of GJ 436b. The outer gaseous envelope of this planet might harbor
a variety of structures in the deep interior \citep[e.g.,][]{Adams2008}. We also apply the theory
to two models of Jupiter, whose core could have a well-defined outer
surface or be diluted due to
erosion \citep[e.g.,][]{Guillot2004}.
We conclude with a discussion of the observability and meaningfulness
of using forthcoming measurements of $k_2$ in the investigation of the interior structure 
of extrasolar planets.

\section{Potential method for calculating the fluid Love numbr $k_{n}$}\label{Sec__PotentialMethod}
The computation of the fluid Love numbers 
requires the solution of the equation \citep[e.g.,][]{Gavrilov1975}

\begin{equation}
    T''_{n}\left(r\right)+\frac{2}{r}T'_{n}\left(r\right)+\left[\frac{4\pi G\rho'\left(r\right)}{V'\left(r\right)}-\frac{n\left(n+1\right)}{r^2}\right]T_n\left(r\right)=0,
    \label{Eq__PotentialEquation}
\end{equation}
where $r,V$, and $\rho$ are the radial coordinate, 
the gravitational potential, and the density of the unperturbed body 
(i.e., spherically symmetric).
A (double) prime indicates (double) derivation with respect to $r$, and $V'(r)=-g(r)$
is the gravitational acceleration.
The function $T$ has the dimensions of a potential and
describes the total change in potential according to
\begin{equation}
    V^{\rm p}+W = \frac{1}{Rg_0}T\left(\frac{R}{r}\right)^{n}W,
\end{equation}
where the perturbation-inducing potential $W$
generates the additional potential $V^{\rm p}$
within the body. 
The gravitational acceleration at the surface is $g_0.$
The perturbed potential is proportional to 
the perturbation-inducing potential 
and at the surface of the deformed body 
($r=R$), the relation is
\begin{equation}
    V^{\rm p}_{n}\left(R\right) = k_{n}W_{n}\left(R\right),
\end{equation}
where the subscript $n$ indicates the degree in the harmonic expansion of $W$. 
From the two equations above, the Love number $k_{n}$ is obtained as 
\begin{equation}
    k_{n}=\frac{T_{n}\left(R\right)}{Rg_0}-1.
    \label{Eq__kn}
\end{equation}

\subsection{Boundary and interface conditions}
At the center, the solution for $T$ must be finite.
At internal density discontinuities of radius $r_{i}$,
both $T$ and its
derivative must be continuous.
This requirement corresponds to 
\begin{eqnarray}
    T_{n}\left(r^-_{i}\right)&=&T_{n}\left(r^+_{i}\right),
    \label{Eq__InternalBC_T}\\
    T'_{n}\left(r^-_{i}\right)&=&T'_{n}\left(r^+_{i}\right) 
    + \frac{4\pi G}{V'\left(r_{i}\right)}
    \left[\rho\left(r^-_{i}\right) - \rho\left(r^+_{i}\right)\right]
    T_{n}\left(r_{i}\right),
    \label{Eq__InternalBC_Tprime}
\end{eqnarray}
where a plus (minus) indicates that the variable is evaluated 
right above (below) the corresponding radius. 
At the surface $\left(r=R\right)$ , the continuity condition is 
\citep{Zharkov1985}
\begin{equation}
    T'\left(R^-\right)=-\frac{\left(n+1\right)}{R}T\left(R^-\right)+ 
    4\pi G \rho\left(R\right)H\left(R\right) + \left(2n+1\right)g_0,  
    \label{Eq__BCSurface}
\end{equation}
where the function $H$ is in relation with $T$ and the 
fluid radial displacement Love number $h_{n}$ through
\begin{eqnarray}
    H&=&-\frac{T}{V'},\\
    h_{n}&=&\frac{H\left(R\right)}{R}.
\end{eqnarray}

\subsection{Solution for a homogeneous sphere}\label{Sec__Homogeneous}
\begin{figure}[h!]
\centering
\includegraphics[width=.4\textwidth]{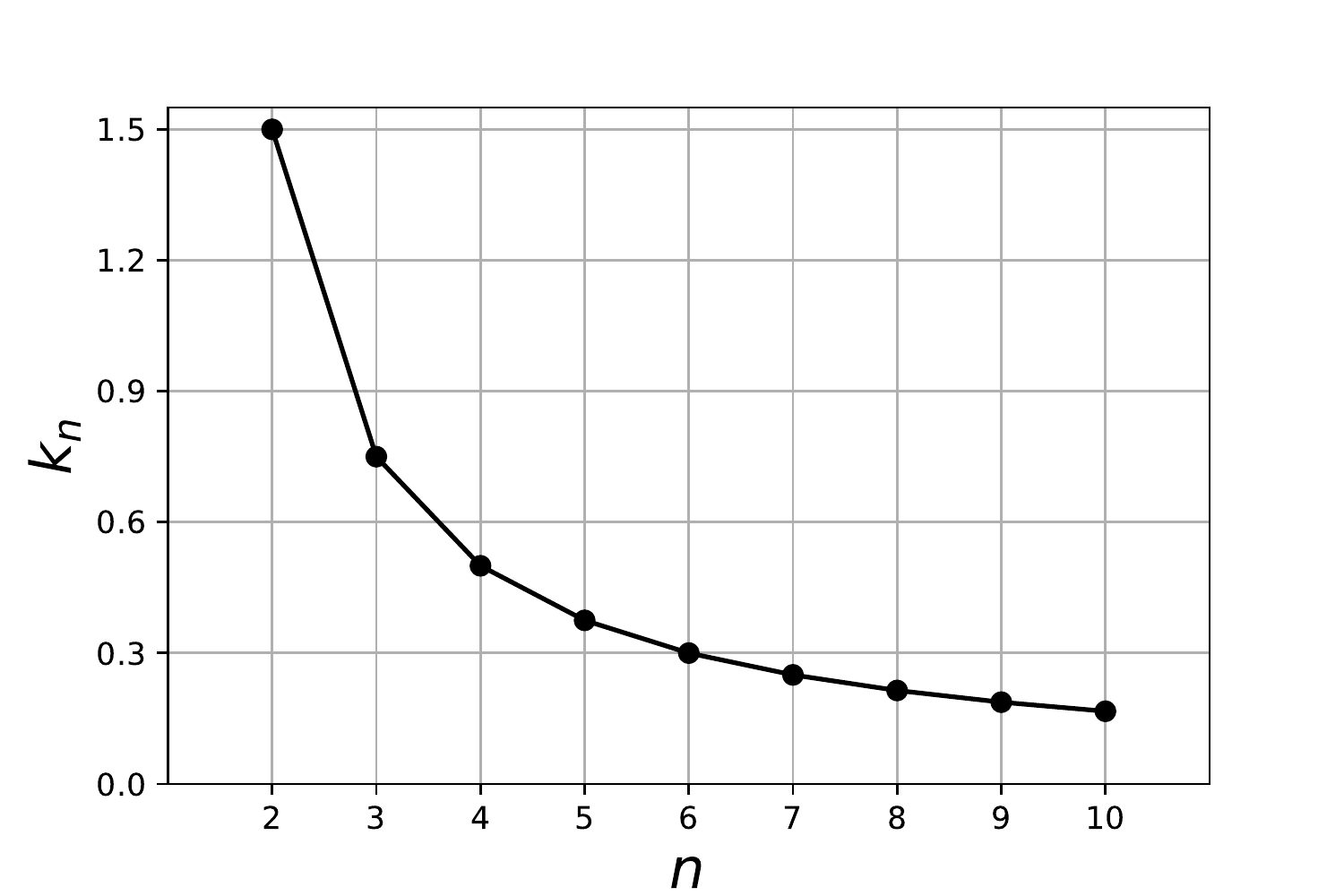}
\caption{Fluid Love number $k$ as a function of the degree $n$ for a homogeneous sphere.}
\label{fig__knHomogeneous}
\end{figure}

If the body has constant density, $\rho'=0,$ and equation (\ref{Eq__PotentialEquation}) 
reduces to an Euler-Cauchy equation.
Using the trial solution $T(r)=r^{c}$ results in
the characteristic equation
\begin{equation}
    c^2+c-n(n+1)=0.
    \label{Eq__CharacteristicEquation}
\end{equation}
Since the discriminant is positive ($n$ is an integer larger than 0) and the 
product of the two solutions is negative, there are two real solutions
with opposite signs for the exponent $c$, 
namely $c_1=n$ and $c_2=-(n+1)$.
The general solution is then the linear combination
\begin{equation}
    T\left(r\right)=Ar^{c_1}+ Br^{c_2},
    \label{Eq__TgeneralSolution}
\end{equation}
where $A$ and $B$ are unknown constants of integration.
The requirement that the function $T$ is finite at the center 
implies that $B=0.$ 
To determine the second integration constant $A,$ the application of the
surface boundary condition from Eq. (\ref{Eq__BCSurface}) gives
\begin{equation}
    T_{n}(r)=\frac{R(2n+1)g_0}{(2n-2)}\left(\frac{r}{R}\right)^{n}.
\end{equation}
The Love number $k_{n}$ of order $n$ is, from Eq. (\ref{Eq__kn}),
\begin{equation}
    k_{n} = \frac{3}{2n-2}.\label{Eq__knHomogeneous}
\end{equation}
The value of $k_{n}$ as a function of $n$ is shown in Figure
\ref{fig__knHomogeneous}.
For $n=2$ the value of $k_{n}$ is 1.5, which represents the limit
corresponding to a normalized moment of inertia of 0.4.

\subsection{Surface deformation}\label{Sec__SurfaceDeformation}
With the value of the fluid Love number $k_{n}$,
the shape of the equipotential surface, which corresponds to the 
physical surface in the fluid limit, can be evaluated from the
value of $h_{n}=1+k_{n}$ and the knowledge of
the perturbing potential (section \ref{Sec__Intro}).  
From the degree-2 term of the tide-inducing potential $W$ in equation
(\ref{Eq__TideInducingPotential}), 
the tidally induced radial deformation at
the surface, $u^{\rm Tidal}_{R}$, can be written as
\begin{eqnarray}
    u^{\rm Tidal}_{R}\left(\psi\right) &=& \frac{h_2}{g_0}W_{2}\left(R,\psi\right),\\
    &=& h_2R\left(\frac{M_{\rm S}}{M}\right)\left(\frac{R}{d}\right)^3P_2\left(\cos\psi\right),
    \label{Eq__TidalRadialDeformation}
\end{eqnarray}
where $M_{\rm S}$, $M$, and $d$ are the stellar (i.e., perturber) mass,
the planetary mass, and
the distance of the perturber.
The shape is rotationally symmetric with respect to the line 
connecting the center of mass of the body with the perturber.
Equation (\ref{Eq__TidalRadialDeformation}) shows that the larger,
less massive, and the closer to the perturber the body, the larger its surface radial deformation (since $u^{\rm Tidal}_{R}\propto\, R^4/M/d^3$).
We note that in general the distance to perturber $d$ varies, being constant 
and equal to the semimajor axis of the orbit  only for circular orbits.
Thus, the shape of the planet, which is assumed to respond as fluid, continually
evolves as the distance $d$ varies.

The rotational potential can be written as a
degree-2 harmonic term as \citep[e.g.,][]{Murray1999}
\begin{eqnarray}
Z\left(r,\theta\right) = \frac{1}{3}\omega^2r^2\left[P_2\left(\cos\theta\right)-1\right],
\end{eqnarray}
where $\omega$ is the angular rotational rate and
$\theta$ is the colatitude measured from the rotation axis. 
The rotational potential is symmetrical with respect to 
the axis of rotation.
As for the tidally induced radial deformation, the 
degree-2 rotationally induced surface radial deformation is
\begin{eqnarray}
    u^{\rm Rotational}_{R}\left(\theta\right) &=& \frac{h_2}{g_0}Z\left(R,\theta\right),\\
    &=&\frac{h_2}{3G}\frac{R^4\omega^2}{M}\left[P_2\left(\cos\theta\right)-1\right].
    \label{Eq__RotationalRadialDeformation}
\end{eqnarray}
Equation (\ref{Eq__RotationalRadialDeformation})
shows that the larger and less massive the body and the faster it rotates, 
the larger its  deformation 
(since  $u^{\rm Rotational}_{R}\propto\, R^4 \omega^2/M$).

In general, the spin axis does not coincide with the axis pointing at the
tidally inducing perturber, and, assuming that the two
perturbations can be added linearly, 
their combined effect
is obtained by expressing them in the same frame
of reference, that is, by 
applying the addition theorem of spherical harmonics.
In the frame of reference of the rotational potential,
\begin{eqnarray}
u_{R}\left(\theta,\phi\right) =  u^{\rm Rotational}_{R}\left(\theta\right) +  u^{\rm Tidal}_{R}\left(\theta,\phi\right)\,\propto\, \frac{R^4h_2}{M}
\label{Eq__LinearCombination}
,\end{eqnarray}
where the dependence on the longitude $\phi$ appears through the rotation of the
reference system where the tidal perturbation is evaluated.
The perturbation thus depends on the fundamental properties of the planet
$R,$ $M,$ and $h_2.$
For degree 2, the combination of rotation and tidal distortion 
corresponds to the sum of a rotationally induced oblate spheroid 
(equation (\ref{Eq__RotationalRadialDeformation})) with a 
tidally induced prolate spheroid
(equation (\ref{Eq__TidalRadialDeformation})).

\subsection{Planet with two constant density layers}
\begin{figure}[t!]
\centering
\includegraphics[width=.5\textwidth]{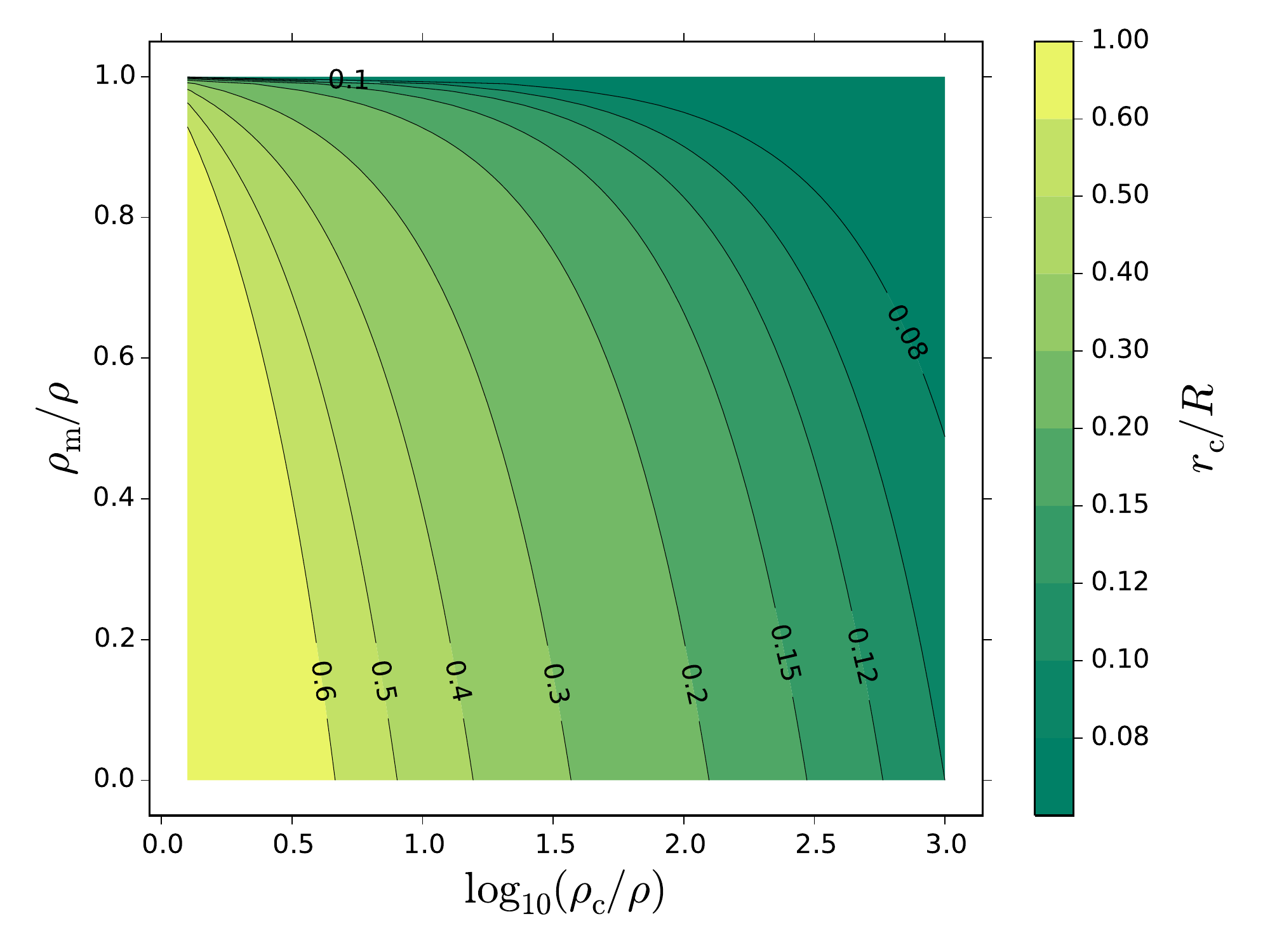}
\caption{Model with two constant density layers. Normalized density of the mantle vs. density of the
core. Colors indicate the normalized radius of the core.
The density of the core is plotted in logarithmic scale.}
\label{fig__2Lmodel}
\end{figure}

This model represents the simplest approximation for a 
differentiated terrestrial (gaseous) planet, where a constant-density mantle
(gaseous envelope) overlies a constant-density 
core.
We indicate with $\rho_{\rm c}$ and $\rho_{\rm m}$ the inner and outer layer densities 
and with $\alpha$ the ratio of the inner layer radius $r_{\rm c}$ to the radius $R$. 
The basic relations for the mean density $\rho$ 
and the mean moment of inertia ${\rm MoI}$ are
\begin{eqnarray}
\rho =\rho_{\rm c}\alpha^3 + \rho_{\rm m}\left(1-\alpha^3\right),\label{Eq__MeanDensity}\\
{\rm MoI} = \frac{2}{5}\left[\frac{\rho_{\rm c}}{\rho}\alpha^5 +  \frac{\rho_{\rm m}}{\rho}\left(1- \alpha^5\right) \right].
\end{eqnarray}
Figure \ref{fig__2Lmodel} illustrates the interplay among
the densities of the two layers (normalized to the mean 
density of the object) and the radius of the core (normalized
to the planetary radius).
The normalized core density is plotted on a log scale,
which includes the case of a small dense core overlaid
by a thick, almost massless, envelope.
For the Love numbers $k_{n}$, in each layer the solution is the same as 
in the homogeneous case, equation (\ref{Eq__TgeneralSolution}).
Continuity across the two layers must be enforced through 
equations (\ref{Eq__InternalBC_T}) and 
(\ref{Eq__InternalBC_Tprime}).
It is then possible to obtain a closed-form solution for $k_{n}.$

\begin{figure}[h!]
\centering
\includegraphics[width=.23\textwidth]{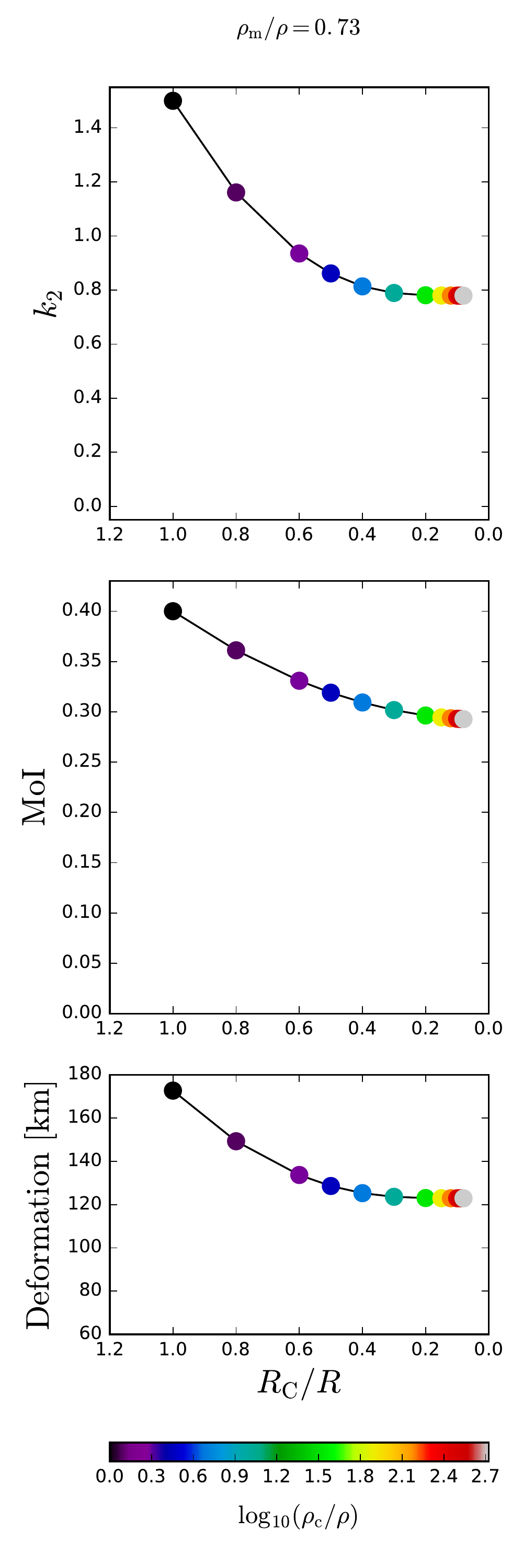}
\includegraphics[width=.23\textwidth]{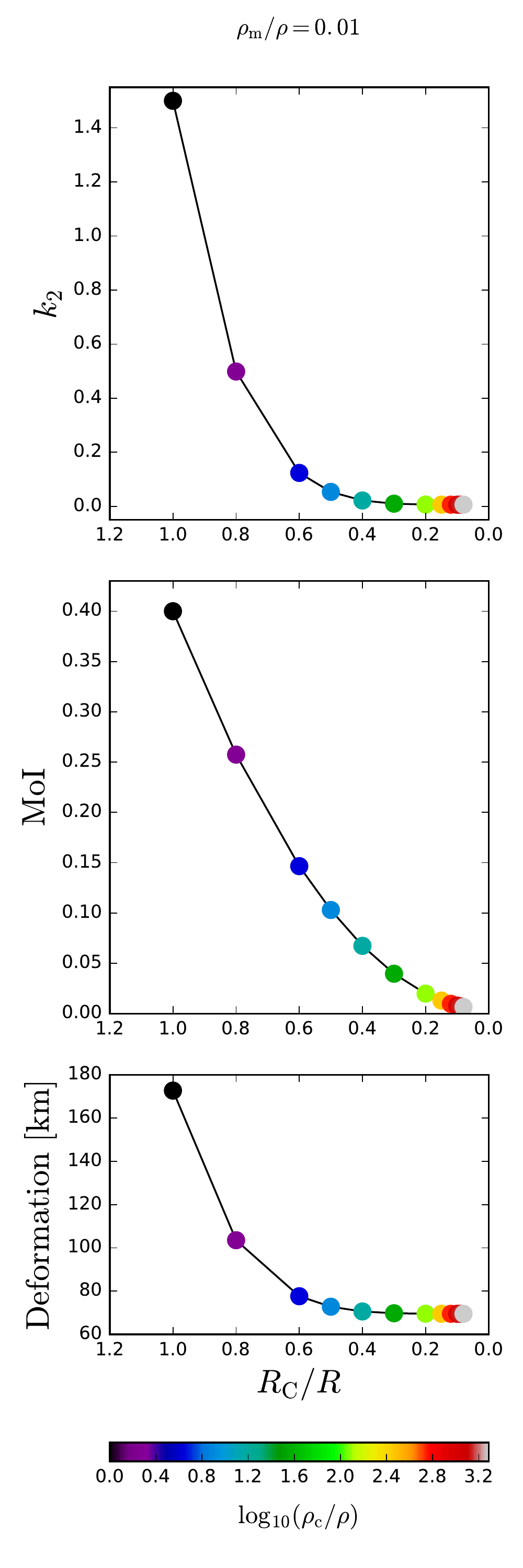}
\caption{ Model of a planet with two constant density layers
with $M=6.55M_{\Earth}$ and $R=2.68R_{\Earth}$.
Values of $k_2$ (first row), normalized
moment of inertia ${\rm MoI}$ (second row), and maximum surface
deformation (third row, assuming $M_{\rm S}=0.157M_{\Sun}$ and
$a=0.0143$ ${\rm AU}$ in equation (\ref{Eq__TidalRadialDeformation})), for  $\rho_{\rm m}/\rho=0.73$  
(left column), and for $\rho_{\rm m}/\rho=0.01$ (right column).
Colors indicate the normalized density of the core.
The color bar is logarithmic.}
\label{fig__GJ1214b}
\end{figure}

The model with two constant density layers allows 
identifying the basic dependences among the 
parameters.
For a planet with a mass and radius similar to
GJ 1214b 
\citep[$M=6.55M_{\Earth}$, $R=2.68M_{\Earth}$,][]{Charbonneau2009},
Figure \ref{fig__GJ1214b} shows the values of $k_{2}$,
the normalized moment of inertia ${\rm MoI}$, and the maximum surface
tidal deformation, which is reached at the sub-stellar and anti-stellar points 
where $P_2(\cos\theta)=1$ in 
equation (\ref{Eq__TidalRadialDeformation}), as 
a function of the core radius, which corresponds,
through equation (\ref{Eq__MeanDensity}), to a core 
density.
For the ratio between the mantle density and
the average density, we use either 0.73, a value similar
to the Earth case, or 0.01, corresponding to a model where
mass is mostly concentrated in a high-density core.
From an inspection of the figure, we note the following:
\begin{itemize}
    \item [1.] The smaller (and thus, the denser) the core, the smaller both 
    $k_2$ and ${\rm MoI}$, consistent with the interpretation of the fluid 
    Love number as a measure of the concentration of mass 
    \citep[e.g.,][]{Kramm2011}.
    \item [2.] A body where most of the mass is concentrated in the core, as    the case for $\rho_{\rm m}/\rho=0.01$ illustrates, has a value of $k_2$ that rapidly
    approaches 0 as the core decreases in size 
    (specifically, for $r_{\rm C}/R=0.08$, $k_2 = 6\times10^{-3}$).
    \item [3.] When $k_2$ approaches 0, this does not imply that the body
    is not deformed, since the deformation depends on $h_2=k_2+1.$ 
    The limiting case of $k_2=0$ (all the mass in a point core overlaid by
    a massless envelope of radius $R$) simply corresponds to $h_2=1$ in 
    equation (\ref{Eq__TidalRadialDeformation}).
    \item [4.] A comparison of the two cases
    in Figure \ref{fig__GJ1214b},
    which only differ in the distribution of the mass in the interior, 
    shows that the three quantities plotted, $k_2$, ${\rm MoI}$, and the
    deformation, decrease with increasing concentration of the mass.
\end{itemize}

\section{Matrix-propagator approach to the calculation of $k_{n}$}\label{Sec__MatrixPropagator}
With $\rho'\left(r\right)=0,$ Equation (\ref{Eq__PotentialEquation}) can be recast as a system
of two first-order differential equations by introducing the
function $P=dT/dr$.
Before writing the system in matrix form, the variables $T$ and
$P$ are non-dimensionalized as follows \citep[e.g.,][]{Wolf1994}:
\begin{eqnarray}
    T&=&\frac{GM}{R}\mbf{y}_1,\label{Eq__y1}\\
    P&=&\frac{dT}{dr}=\frac{GM}{R^2}\left(\frac{R}{r}\right)\mbf{y}_2, \label{Eq__y2}
\end{eqnarray}
where $\mbf{y}_{i}$ are non-dimensional variables. 
The derivatives of $T$ and $P$ with respect to $r$ are
\begin{equation}
    \frac{dT}{dr}=\frac{GM}{R^2}\frac{d\mbf{y}_1}{ds};\;\;\;\;\;\;\frac{dP}{dr}=\frac{GM}{R^3}\left(\frac{1}{s}\frac{d\mbf{y}_2}{ds}-\frac{\mbf{y}_2}{s^2}\right),
\end{equation}
where $s=r/R$ is the non-dimensional radius.
The non-dimensional system of equations is
\begin{eqnarray}
   \frac{d\mbf{y}_{1}}{ds}&=&\frac{\mbf{y}_2}{s},\\
   \frac{d\mbf{y}_2}{ds}&=&\frac{n\left(n+1\right)}{s}\mbf{y}_1-\frac{1}{s}\mbf{y}_2.
\end{eqnarray}
The system can 
be written in matrix form as
\begin{eqnarray}
   \frac{d}{ds}\left[\mbf{y}\right]=
   \frac{d}{ds}\left[\begin{array}{c} \mbf{y}_1 \\ \mbf{y}_2 \end{array} \right] = 
   \left[\begin{array}{cc}
        0 & \frac{1}{s} \\
        \frac{n(n+1)}{s} & -\frac{1}{s}
   \end{array}\right]
   \left[\begin{array}{c} \mbf{y}_1 \\ \mbf{y}_2 \end{array} \right].
   \label{Eq__MatrixSystem}
\end{eqnarray}
Assuming power solutions of the form $\mbf{y}_{i}=s^{k^{(j)}}\mrm{y}_{i}^{(j)}$, 
with $\mrm{y}_{i}^{(j)}$ a constant, the system (\ref{Eq__MatrixSystem}) becomes
\begin{eqnarray}
   \left[\begin{array}{cc}
    k^{(j)} & -1 \\
    -n(n+1) & 1+k^{(j)} 
   \end{array}\right]
   \left[\begin{array}{c}
       \mrm{y_1}^{(j)} \\
       \mrm{y_2}^{(j)} 
   \end{array}\right]
   = 0,
\end{eqnarray}
which admits non-trivial solutions only if the determinant vanishes, that is, if
\begin{equation}
    {k^{(j)}}^2 + k^{(j)} -n\left(n+1\right) = 0.
\end{equation}
The last equation corresponds to the characteristic equation 
(\ref{Eq__CharacteristicEquation}), with solutions
\begin{equation}
    k^{(1)} = n, \; k^{(2)} = -\left(n+1\right)
.\end{equation}
Thus, the general solution for $\mbf{y}_{i}$ is the linear combination
\begin{equation}
    \mbf{y}_{i} = \sum_{{j}=1}^2C_{(j)}s^{k(j)}\mrm{y}_i^{(j)},
\end{equation}
where the solution vectors $\mrm{y}_i^{(j)}$ are 
\begin{equation}
      \mrm{y}_i^{(j)}  = \left[\begin{array}{c} 1 \\ k^{(j)} \end{array} \right]. 
\end{equation}
The general solution in matrix form is then\begin{equation}
    \mbf{y}=\mbf{PC},
    \label{Eq__MatrixGeneralSolution}
\end{equation}
where
\begin{eqnarray}
  \mbf{P}&=& \left[\begin{array}{cc}
    s^{n}  & s^{-(n+1)} \\
    ns^{n} & -(n+1)s^{-(n+1)} 
   \end{array}\right],\label{Eq__MatrixP}\\
   \mbf{C}&=&\left[C_1\;\;C_2\right]^{\mrm{T}}.
\end{eqnarray}
$\mbf{P}$ is the propagator matrix and $\mbf{C}$ is the vector of
constants of integration.

\subsection{Internal boundary conditions in matrix form}
Indicating with $r_{i}$ the radius of a density discontinuity,
the two internal boundary conditions, equations
(\ref{Eq__InternalBC_T}) and (\ref{Eq__InternalBC_Tprime}),
can be expressed in matrix form as
\begin{equation}
   \left[\mbf{y}^+\right]_{r_{i}}=
   \left[\begin{array}{c} \mbf{y}_1^+ \\ \mbf{y}_2^+ \end{array} \right]_{r_{i}} = 
   \left[\begin{array}{cc}
        1 & 0 \\
        -\frac{4\pi G r_{{i}}\Delta\rho }{g\left(r_{i}\right)} & 1 
   \end{array}\right]
   \left[\begin{array}{c} \mbf{y}_1^- \\ \mbf{y}_2^- \end{array} \right]_{r_{i}},
   \label{Eq__MatrixInterface}
\end{equation}
where a plus (minus) indicates that the variable is evaluated 
right above (below) the boundary. 
The density difference $\Delta\rho=\rho(r_{i}^-) - \rho(r_{i}^+)$
is positive for gravitationally stable planets, for which density increases
downward.

\subsection{Propagator}\label{Sec__Propagator}
The planetary models used in this study are made of a series of constant property layers.
Eq. (\ref{Eq__MatrixGeneralSolution}) provides the general solution in each layer.
The interface conditions between layers in matrix form corresponds to the square matrix
in Eq. (\ref{Eq__MatrixInterface}):
\begin{equation}
    \mbf{B}\left(r_{i},\Delta\rho_i\right)=
   \left[\begin{array}{cc}
        1 & 0 \\
        -\frac{4\pi G r_{{i}}\Delta\rho }{g\left(r_{i}\right)} & 1 
   \end{array}\right].
\end{equation}
\begin{figure}[t!]
\begin{center}
\includegraphics[width=.4\textwidth]{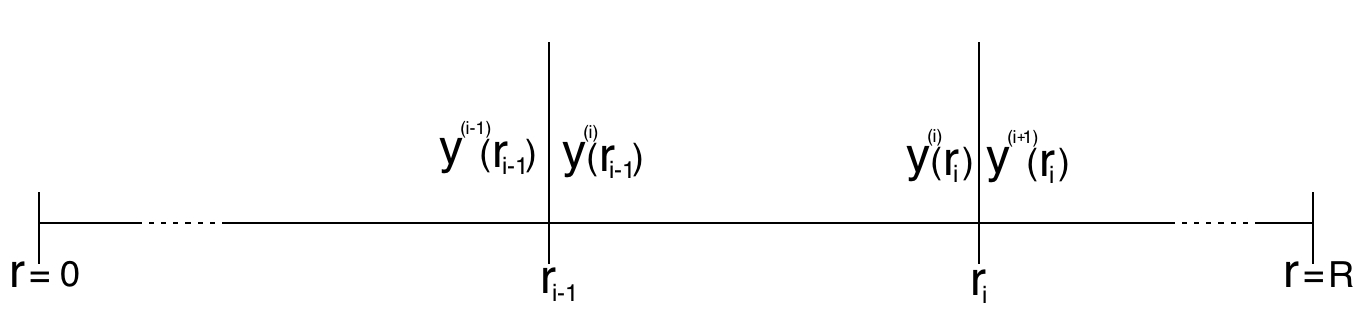}
\caption{\small{Schematic structure of the vector of solution $\mathtt{y}$ and radius $r$ for a planetary model. The center of the planet is at $r=0,$ the surface at $r=R.$
Each homogeneous layer is indicated by the index corresponding to the outer boundary of the layer.
Similar indexing applies to the non-dimensional radius $s$.}}
\label{Fig__Viscel_Structure}
\end{center}
\end{figure}
With reference to the interface at $r=r_{i-1}$ in Figure \ref{Fig__Viscel_Structure}, the two solutions to be matched are
\begin{eqnarray}
\mbf{y}^{(i)}(s_{i-1})  &=&\mbf{P}_i(s_{i-1})\mbf{C}^{(i)},\label{Eq_Visc_y4}\\
\mbf{y}^{(i-1)}(s_{i-1})&=&\mbf{P}_{i-1}(s_{i-1})\mbf{C}^{(i-1)}.
\end{eqnarray}
The interface matrix $\mbf{B}(r_{i-1})$ provides the connection between the two solutions:
\begin{equation}
\mbf{y}^{(i)}(s_{i-1})=\mbf{B}(r_{i-1},\Delta\rho_{i-1})\mbf{y}^{(i-1)}(s_{i-1}).
\end{equation}
With the last three expressions, the vector of constants $\mbf{C}^{(i)}$ can be expressed in terms of the vector of constants $\mbf{C}^{(i-1)}:$
\begin{equation}
\mbf{C}^{(i)}=\left[\mbf{P}_{i}(s_{i-1})\right]^{-1}\mbf{B}(r_{i-1},\Delta\rho_{i-1})\mbf{P}_{i-1}(s_{i-1})\mbf{C}^{(i-1)}\label{Eq_Visc_C4}.
\end{equation}
From eq. (\ref{Eq_Visc_y4}) and (\ref{Eq_Visc_C4}), the solution for $\mbf{y}^{(i)}$ at $r=r_{i}$ is
\begin{equation}
\mbf{y}^{(i)}(s_{i})=\mbf{P}_{i}(s_{i})\left[\mbf{P}_{i}(s_{i-1})\right]^{-1}\mbf{B}(r_{i-1},\Delta\rho_{i-1})\mbf{P}_{i-1}(s_{i-1})\mbf{C}^{(i-1)},
\end{equation}
and the vector of integration constant $\mbf{C}^{(i)}$ no longer appears. 
By extending this approach to each layer, the solution $\mbf{y}^{(N)}$ at $r_N$ can then be written as the product of a sequence of terms and the vector of constants at the center of the planet:
\begin{eqnarray}
\mbf{y}^{(N)}(s_N)&=&\left\{\prod^{N}_{k=2}\mbf{P}_k(s_k)\left[\mbf{P}_k(s_{k-1})\right]^{-1}\mbf{B}(r_{k-1},\Delta\rho_{k-1})\right\}\\
&&\cdot\mbf{P}_1(s_1)\mbf{C}^{(1)}.
\label{Eq__Propagator}
\end{eqnarray}

\subsection{Simplification of the propagator matrix product}\label{Sec__ComputationalTechnicalities}
To compute the product of $\mbf{P}$ with its inverse in Eq. (\ref{Eq__Propagator}), 
it is convenient to write, using Eq. (\ref{Eq__MatrixP}),
\begin{equation}
  \mbf{P}(r)=
  \left[\begin{array}{cc}
    1  & 1 \\
    n  & -(n+1)
   \end{array}\right]\left[\begin{array}{cc}
    r^{n}  & 0 \\
    0 & r^{-(n+1)} 
   \end{array}\right],
\end{equation}
so that
\begin{equation}
  \left[\mbf{P}(r)\right]^{-1}=-\frac{r}{(2n+1)}
   \left[\begin{array}{cc}
    r^{-(n+1)}  & 0 \\
    0 & r^{n} 
   \end{array}\right]
  \left[\begin{array}{cc}
    -(n+1)  & -1 \\
    -n  & 1 
   \end{array}\right].
\end{equation}
With the last two expressions
\begin{eqnarray}
\mbf{P}(r_j)\left[\mbf{P}(r_{j-1})\right]^{-1}=
\mbf{D}(n)\mbf{X}(n,r_j,r_{j-1})\left[\mbf{D}(n)\right]^{-1},
\end{eqnarray}
where
\begin{eqnarray}
\mbf{D}(n)&=&\left[\begin{array}{cc}
    1  & 1 \\
    n  & -(n+1)
   \end{array}\right],\\
   \mbf{X}(n,r_j,r_{j-1})&=& \left[\begin{array}{cc}
    \left(\frac{r_j}{r_{j-1}}\right)^n  & 0 \\
    0  & \left(\frac{r_j}{r_{j-1}}\right)^{-(n+1)}
   \end{array}\right].
\end{eqnarray}

\subsection{Solution}
With the results of Sections \ref{Sec__Propagator} and \ref{Sec__ComputationalTechnicalities},
we can write the general solution at the surface as
\begin{eqnarray}
    \mbf{y}(R)=\mbf{M}\mbf{y}(0),
\end{eqnarray}
where the matrix $\mbf{M}$ is completely defined by the interior model, that is, by the 
density as a function of radius.
Explicitly,
\begin{eqnarray}
      \left[\begin{array}{c} \mbf{y}_1(R) \\ \mbf{y}_2(R) \end{array} \right] =
        \left[\begin{array}{cc}
            M_{11}  & M_{12} \\
            M_{21}  & M_{22}
       \end{array}\right]
      \left[\begin{array}{c} \mbf{y}_1(0) \\ \mbf{y}_2(0) \end{array} \right],
      \label{Eq__FullSolution}
\end{eqnarray}
where $\mbf{y}_2(0)=0$, according to the definition of $P$ in 
Eq. (\ref{Eq__y2}). 
The surface boundary conditions, Eq. (\ref{Eq__BCSurface}),
in terms of $\mbf{y}_1$ and $\mbf{y}_2$ read
\begin{eqnarray}
    \mbf{y}_2(R) = -(n+1)\mbf{y}_1(R)+\frac{4\pi G\rho(R)R}{g_0}\mbf{y}_1(R) + (2n+1).
    \label{Eq__BCSurfacey1y2}
\end{eqnarray}
Using Eq. (\ref{Eq__BCSurfacey1y2}) and Eq. (\ref{Eq__FullSolution})
\begin{eqnarray}
        \left[\begin{array}{cc}
            M_{11}  & -1 \\
            M_{21}  & (n+1) -\frac{4\pi G\rho(R)R}{g_0} 
       \end{array}\right]
      \left[\begin{array}{c} \mbf{y}_1(0) \\ \mbf{y}_1(R) \end{array} \right]=
      \left[\begin{array}{c} 0 \\ (2n+1) \end{array} \right],
\end{eqnarray}
whose inversion provides the boundary conditions of the problem.
From the value of $\mbf{y}_1(R)$, through Eq. (\ref{Eq__y1}) and (\ref{Eq__kn}) the 
Love number for any $n$ can be determined.

\subsection{Sub-layering}\label{Sec__Layering}
\begin{figure}[t!]
\centering
\includegraphics[width=.4\textwidth]{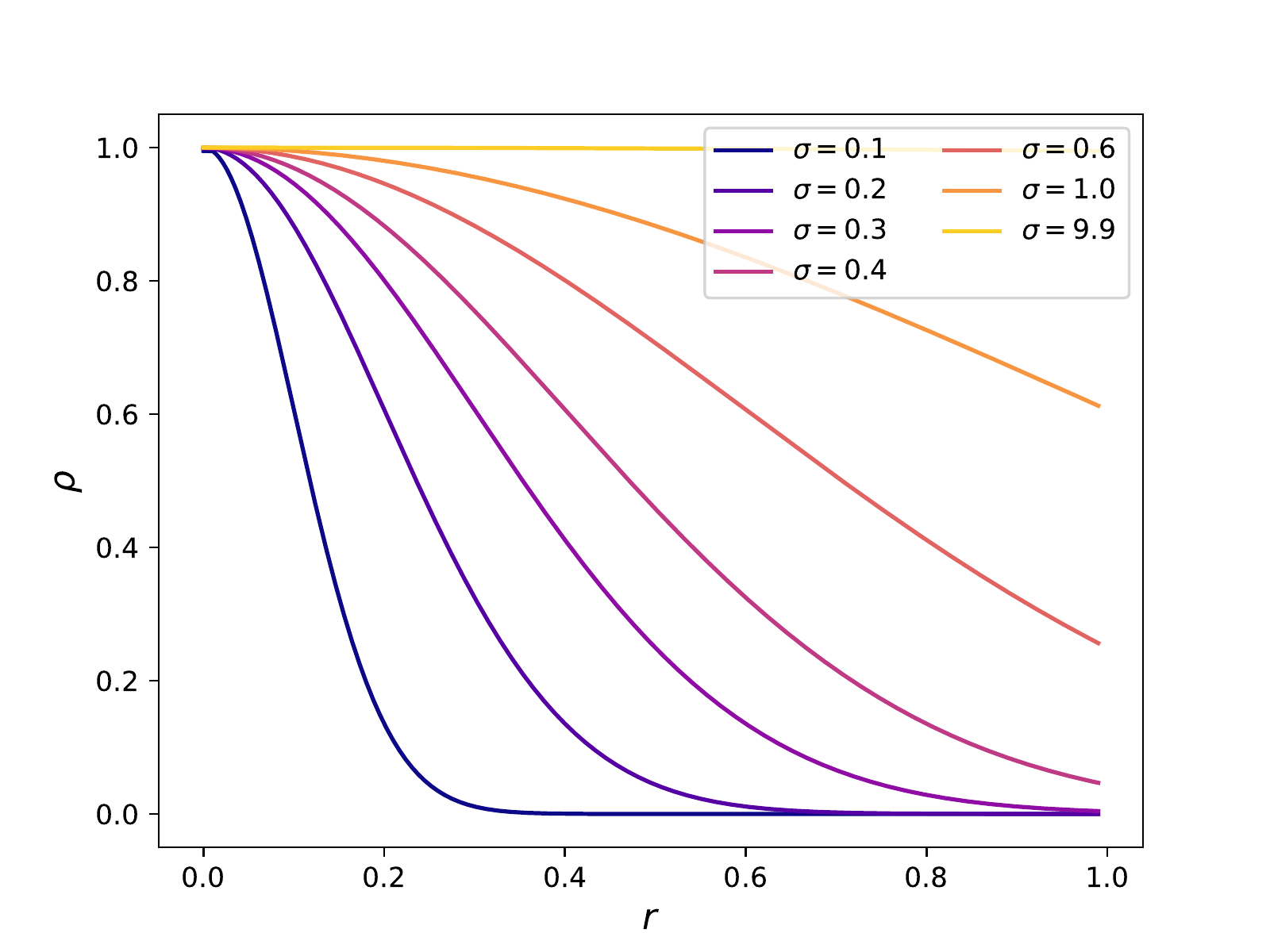}
\caption{Set of normalized interior density profiles obtained using Gaussian curves with different
standard deviations, as indicated in the legend. The yellow curve represents a homogeneous
planet, and the blue curve shows a planet with a high concentration of mass in the interior.}
\label{Fig__DensityProfiles}
\end{figure}
\begin{figure}[t!]
\centering
\includegraphics[width=.45\textwidth]{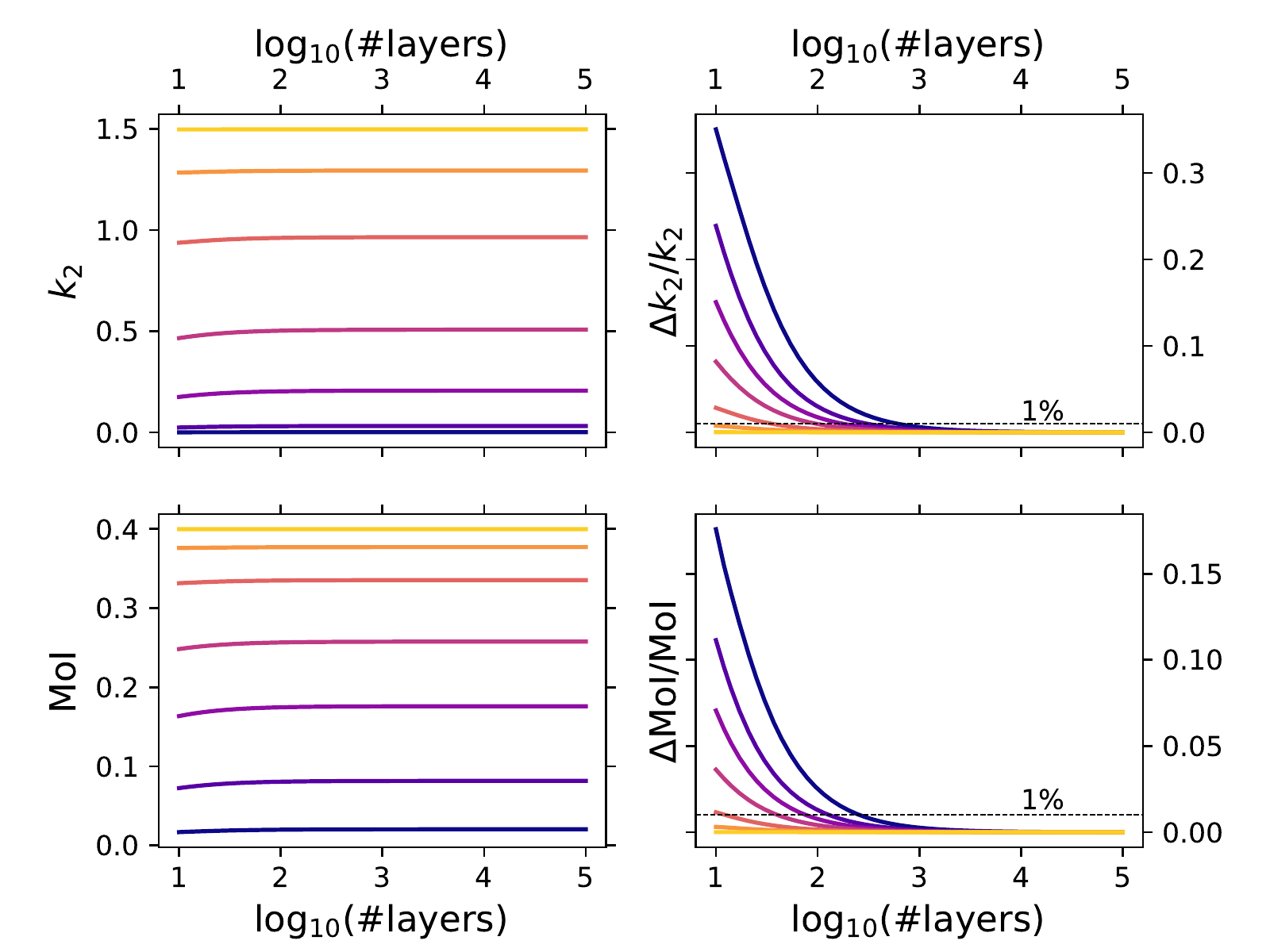}
\caption{For the models of Figure \ref{Fig__DensityProfiles}, 
the left column shows values of the Love number $k_2$
and of the normalized moment of inertia (MoI) as a function of the 
number of layers used to discretize the density profile.
For a given number of layers, the right column shows the relative variation
with respect to the value for $10^5$ layers.}
\label{Fig__k2MoI}
\end{figure}
\begin{figure}[t!]
\centering
\includegraphics[width=.4\textwidth]{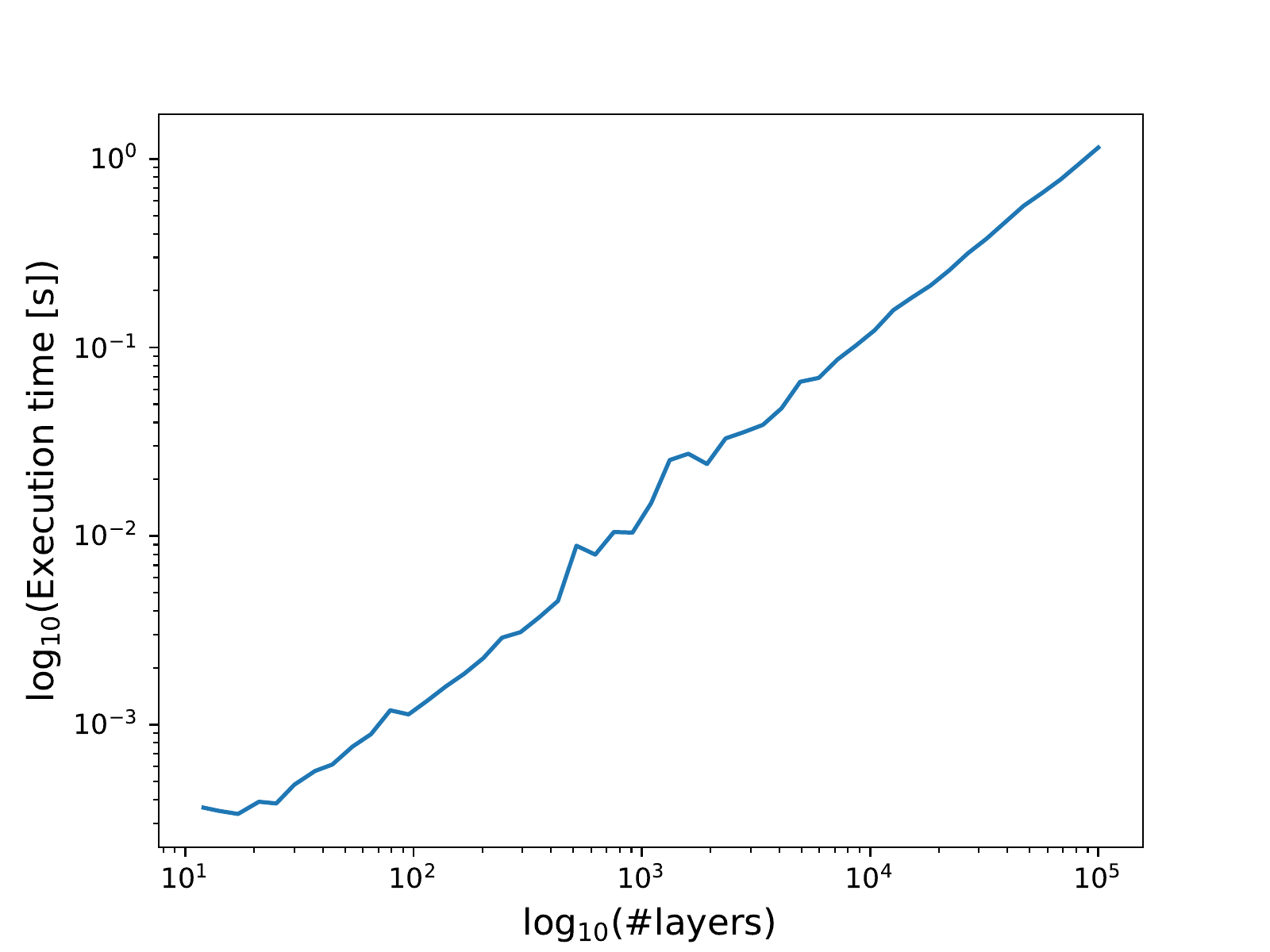}
\caption{Speed of computation vs. number of layers in the calculation of the Love number $k_2$
for a single-density profile. The curve is similar for $n=2,...,8$. 
A 2.5 GHz Intel Core i7 processor was used.}
\label{Fig__k2ComputationTime}
\end{figure}

Given the discretized nature of the matrix-propagator approach,
the solution does depend on the assumed number of layers,
and a suitable choice should be made to ensure an accurate
result. 
To illustrate the dependence on the layering, 
we generated seven interior models where the 
density profile is obtained as the portion of a Gaussian
curve with zero mean and standard deviation $\sigma:$
\begin{equation}
        \rho\left(r\right) = \exp\left[{-\frac{r^2}{2 \sigma^2}}\right].
        \label{Eq__Gaussian}
\end{equation}
By using only values of $r$ between 0 and 1, 
these models cover a range of concentration of 
mass in the interior, from an almost
homogeneous distribution ($\sigma = 9.9$) 
to a concentrate one ($\sigma = 0.1$). 
Their density profiles 
are plotted in Figure \ref{Fig__DensityProfiles}.
We computed the moment of inertia and the Love 
number $k_2$ as a function of the number
of sub-layers used to to discretize the analytical 
profile expressed in equation (\ref{Eq__Gaussian}),
and the results are shown in Figure \ref{Fig__k2MoI}.
Both $k_2$ and the MoI converge to
a value that represents the ``continuous'' limit. This is the
limit of an infinite number of sub-layers with zero thickness.
In the right column of Figure \ref{Fig__k2MoI} 
we plot the relative variation
of the values of $k_2$ and the MoI with respect to the limit
value, which we take as the one corresponding to 
 $10^5$ layers.
The more homogeneous a planet, 
the smaller the density variations as a 
function of the radius,
and correspondingly, the smaller the number of layers 
necessary to retrieve an accurate result,
as the comparison of the blue and yellow curves indicates.
In all cases, using a number of sub-layers in excess of about $10^3$ 
allows the retrieval of 
$k_2$ and the MoI 
with a precision better than 1\%.
The required accuracy of the modeling can
guide the choice of the number of sub-layers to be used.
Figure \ref{Fig__k2ComputationTime} shows that the computation of
$k_2$ for a model with $10^5$ sub-layer  requires approximately 1 second.
When thousands of models are involved, Figures \ref{Fig__k2MoI} 
and \ref{Fig__k2ComputationTime} provide
a guide to the tradeoff between precision and speed.

The results regarding the precision as a function of the 
number of sub-layers do depend on the assumed interior model,
and the Gaussian curves used in this section
represent an example that can be adapted on a case-by-case basis.
However, in computing the Love numbers of the Earth
(section \ref{Sec__Validation}), of planet 
\object{GJ 436b}
(section \ref{Sec__GJ436b}),
and of Jupiter (section \ref{Sec__Jupiter}), 
we found that a number of sub-layers between
$10^3$ and $10^4$ results in a precision of 1\%
or better.

\subsection{Validation}\label{Sec__Validation}
We used three analytical density profiles
for Jupiter, linear, quadratic, and polytropic, as proposed
by \citet{Gavrilov1976}. With these, we 
inferred values for the Love number $k_{n}$ for
$n=2,...7.$
\citet{Gavrilov1976} solved the full equation
(\ref{Eq__PotentialEquation}), while
we set the term proportional 
to $\rho'$ equal to zero
and used $10^3$ sub-layers to reproduce the 
smooth variation of density with radius.
Our results for the Love numbers
match those of \citet{Gavrilov1976},
showing that  
the matrix-propagator method
returns the values obtained with the solution of the
full equation, 
provided an
appropriate number of sub-layers is chosen 
(section \ref{Sec__Layering}).

In order to assess the precision of the method, we tested it using PREM,
the preliminary reference Earth model
\citep{Dziewonski1981}, as tabulated in the file
PREM\_1s.csv of the IRIS database 
\citep{Trabant2012}.
For the crust and upper mantle, we used the PREM data directly
(108 layers for $r>5701$ km).
For each of inner core, outer core, and lower mantle, 
we fit the PREM data with a third-degree polynomial,
which was then used to generate $10^{\rm m}$ sub-layers,
with m=2,...,5 (i.e., we built four Earth models with a total number of 
layers of $3\times10^{\rm m}+108$).
The corresponding values of $k_2$ and moment of inertia
are listed in Table \ref{Tab__PREM}.
Even with a few hundred layers, the results are below 
one per thousand of the reference values.

Finally, we compared the results obtained with 
the matrix-propagator method with those computed with the 
method of the concentric Maclaurin spheroids \citep{Hubbard2013}.
Using the preferred density profile of Jupiter
from \citet{Wahl2016}, we obtain a value of $k_2$ 
that differs by less than 0.1\% from the non-rotating Jupiter
case \citep[Table 4 in][]{Wahl2016}.
The comparison with the rotating Jupiter case cannot be
made in the framework of the Love numbers as defined here
(section \ref{Sec__DiscussionNonLinearity}).

\begin{table}[t!]
\caption{Fluid Love number $k_2$ and normalized moment of inertia of the Earth.}
\centering
\begin{tabular}{ r | c c}
\hline\hline
\# sub-layers & $k_2$ & ${\rm MoI}$ \\
\hline
408       &  0.933691  & 0.330883\\
3108     &   0.933433 &  0.330847\\
30108   &   0.933408 &  0.330843\\
300108 &    0.933406 &  0.330843 \\\hline
Reference & 0.934 & 0.3308  \\
\hline
\end{tabular}
\tablefoot{Values for $k_2$ and moment of inertia calculated using 
a discretized version of PREM (see text for details). The reference values
for $k_2$ and the MoI are from \citet{Lambeck1980} and the NSSDC (nssdc.gsfc.nasa.gov),
respectively.}
\label{Tab__PREM}
\end{table}

\section{Applications}
\subsection{GJ 436b}\label{Sec__GJ436b}
\begin{figure}[h!]
\centering
\includegraphics[width=.5\textwidth]{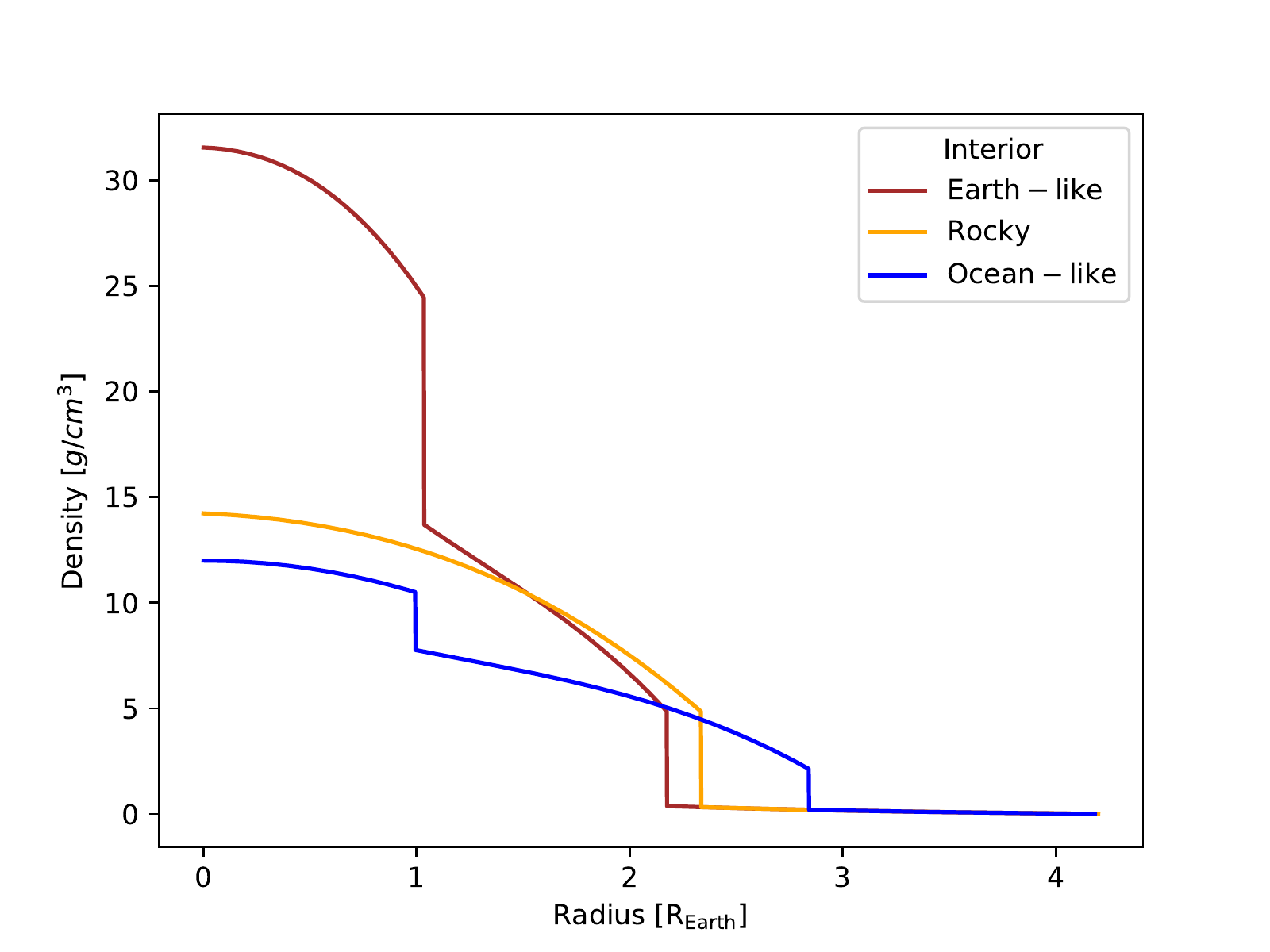}
\includegraphics[width=.5\textwidth]{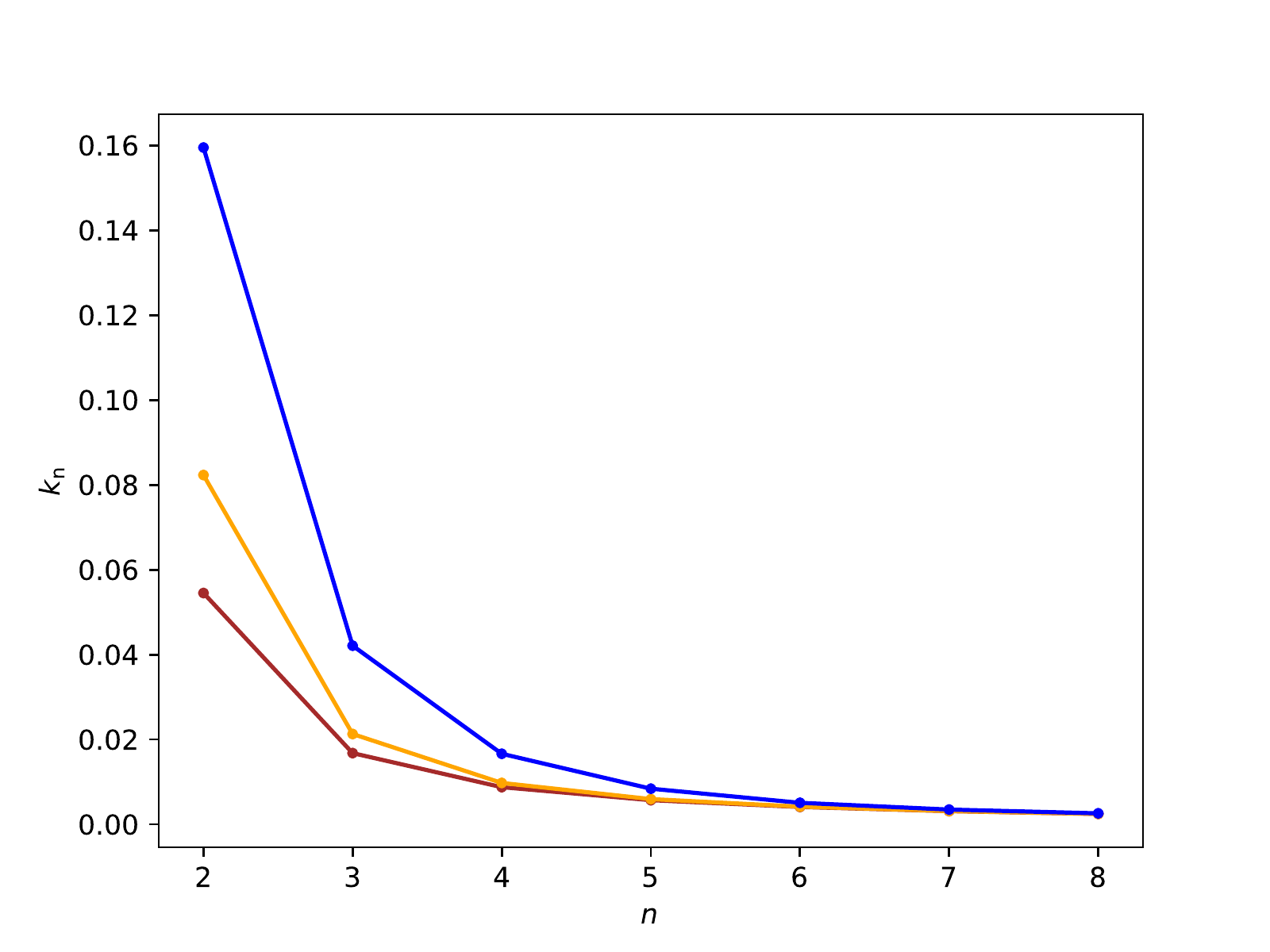}
\caption{(Top) Three possible density profiles for 
\object{GJ 436b}
with the same mass and radius, 21.4 $M_{\Earth}$ and 4.19 $R_{\Earth}$.
A low-density hydrogen or hydrogen/helium envelope surrounds an Earth-like (brown), rocky
(orange), or ocean-like (blue) interior. (Bottom) Corresponding values 
of the Love numbers $k_{n}$.}
\label{Fig__GJ436b}
\end{figure}

The extrasolar planet GJ 436b
is a classical example of the
degeneracy\footnote{The term degeneracy simply indicates
that multiple solutions are possible for a given set of
observations. It has the same meaning as non-uniqueness,
which 
is more commonly used in the geophysical community.} 
of mass-radius relationships 
\citep[e.g.,][]{Adams2008,Nettelmann2010,Kramm2011}.
Here we followed the analysis and nomenclature 
of \citet{Adams2008} and considered
three possible interior structures, where a gaseous 
helium or hydrogen/helium envelope surrounds
an Earth-like, a rocky, or an ocean-like interior.
Using a newly developed interior structure code
\citep{Baumeister2018},
we recalculated the models of \citet{Adams2008} 
to make them compatible
with the most recent values for
the mass \citep[$21.4M_{\Earth}$,][]{Trifonov2018} 
and radius \citep[$4.191R_{\Earth}$,][]{Turner2016} 
of the planet. 
The profiles we obtained fit the 
observed mass and radius 
with a relative error smaller 
than 0.1\%.
To these profiles we applied the matrix-propagator method
of section \ref{Sec__MatrixPropagator}
to compute the Love numbers $k_{n}$ for
$n=2,...,8.$
The density profiles and the corresponding Love numbers
are plotted in Figure \ref{Fig__GJ436b}.
We note that the models considered here do not span the 
entire range of proposed interior structures of the planet
\citep[see, e.g.,][]{Nettelmann2010,Kramm2011}, but
are used as illustration.

The value of $k_{n}$ decreases with increasing $n$, as in the case
of the homogeneous sphere (Figure \ref{fig__knHomogeneous}).
To first order, the absolute
value of $k_2$ is controlled by the
density difference between the center and the surface,
which is a proxy for the degree of mass concentration.
The Earth-like interior model, with a metallic core that reaches 
a density in excess of 30 g/cm$^3$, has $k_2=0.055$, while 
the ocean-like interior model, with a central 
density of about 12 g/cm$^3$, has $k_2=0.160$.
The rocky interior model, with an intermediate central density,
has a value of $k_2=0.082.$
Unlike the moment of inertia, 
the value of $k_2$ tends to rapidly decrease toward zero
as the mass concentration increases
(compare the left panels of Figure \ref{Fig__k2MoI}).
This observation explains why these models, which have
an extended light gaseous envelope and thus 
are very  concentrated,
have all similar low values for $k_2$.

In general, the higher $n$, the closer the
region of the interior that contributes to the 
corresponding $k_{n}$
\citep[e.g.,][]{Gavrilov1976}.
Thus, it is expected that for increasing
$n$ the curves tend to converge, since
these models have a similar gaseous envelope.
This is true in particular for the Earth-like
and rocky models, where the thickness of the
envelope is similar.

The most likely parameter
to be measured for exoplanets is $k_2$ 
\citep{Ragozzine2009,Hellard2018},
and Figure \ref{Fig__GJ436b} shows
that an accurate determination of $k_2$
could more easily distinguish the ocean-like
interior model from the other two.
Although these three models are only illustrative
of the degeneracy of a mass and radius determination
\citep{Adams2008},
the additional modeling  
of the Love number $k_2$ 
shows how 
its potential measurement
would help in breaking the mass-radius degeneracy, at least partly.

\subsection{Jupiter-like hot Jupiter}\label{Sec__Jupiter}

\begin{figure}[h!]
\centering
\includegraphics[width=.5\textwidth]{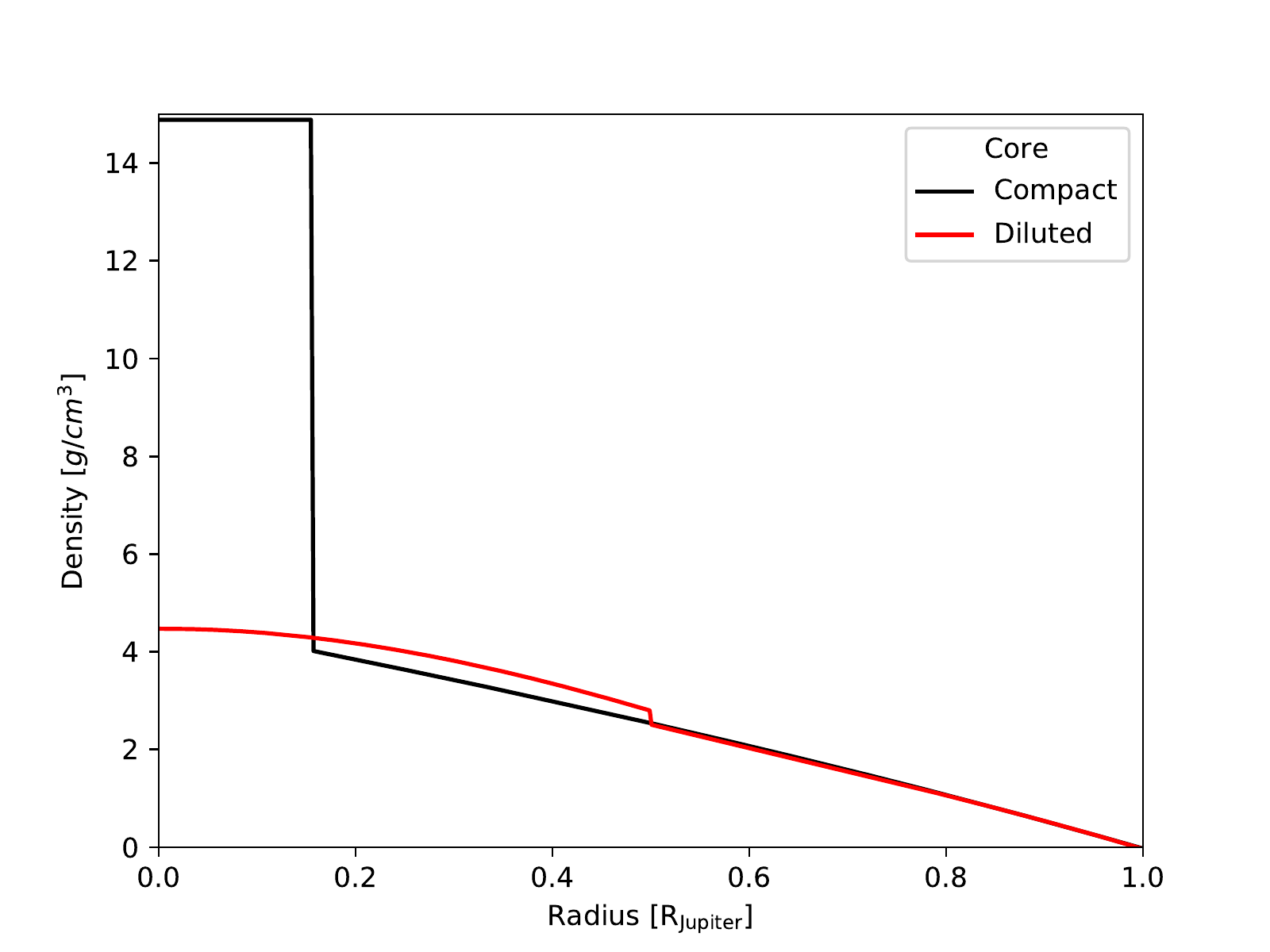}
\includegraphics[width=.5\textwidth]{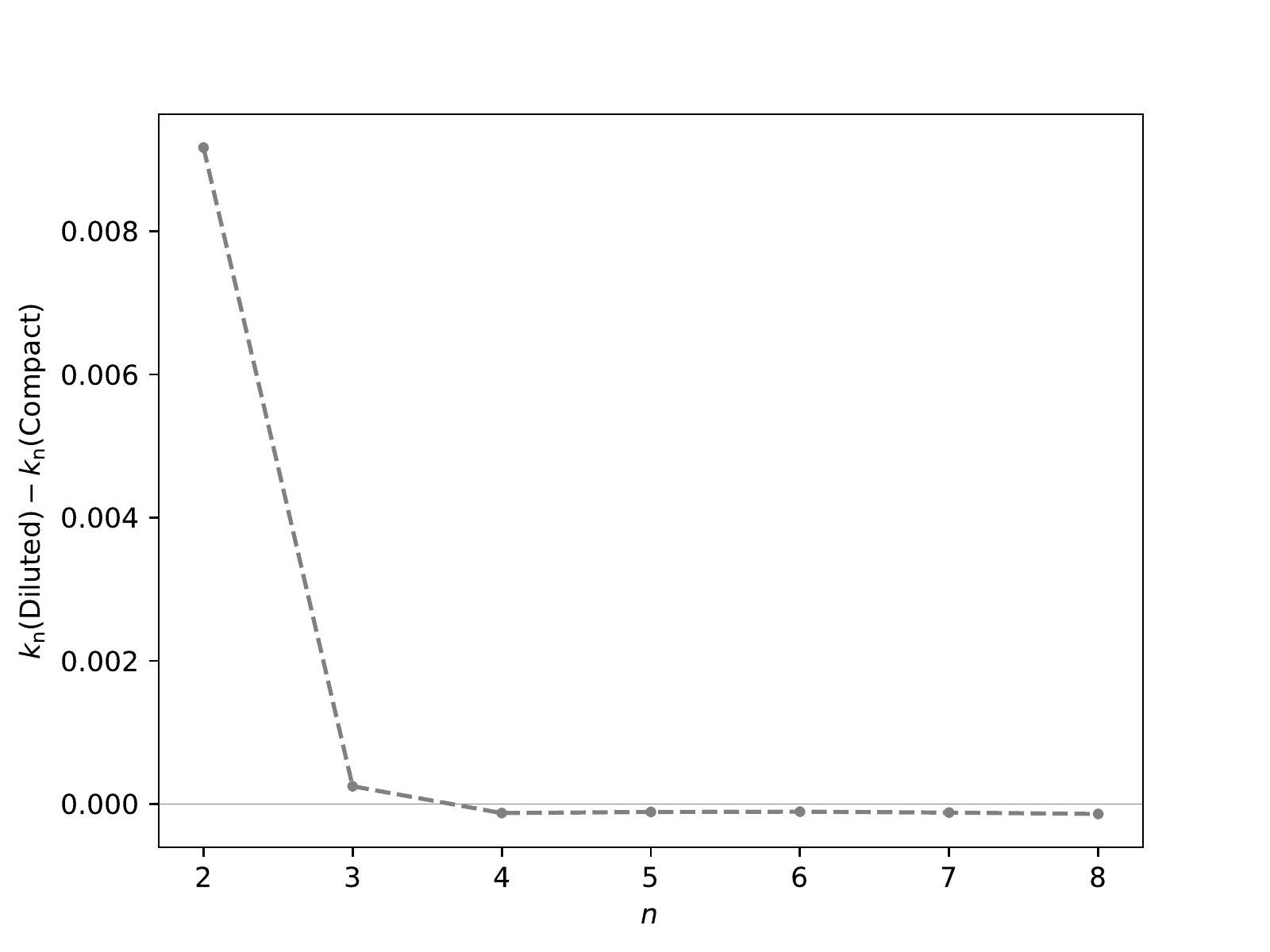}
\caption{(Top) Interior models of Jupiter with
a compact high-density core (black) and a diluted
and extended central region (red). Data are taken from  \citet{Wahl2017}.
(Bottom) Difference in the values of the 
Love numbers $k_{n}$ for the two models.}
\label{Fig__Jupiter}
\end{figure}

Jupiter is a fast rotator, and the calculation of its Love numbers
requires the use of the concentric Maclaurin spheroids method 
(section \ref{Sec__DiscussionNonLinearity}).
Here, we use it as a representative model of hot Jupiters, for which
the linear theory developed above is accurate.
The deep interior of this gas giant is enriched in heavy elements,
which could be segregated into a compact, high-density core or 
diluted into a more
extended, enriched central region \citep[e.g.,][]{Guillot2004}.
We used two density profiles appropriate for these two scenarios
\citep[from][]{Wahl2017} and computed the Love numbers $k_{n}$ for
$n = 2,...,8$.
In Figure \ref{Fig__Jupiter} we plot the profiles and the differences
in the values of $k_{n},$ given that they are quite similar
except for $n=2.$
The higher $n$, the
shallower the region of the interior that mostly contributes
to the corresponding $k_{n}$ \citep[Figure 2b]{Gavrilov1976}.
Thus, given the similarity of the two profiles for $r\gtrsim0.15,$
the Love numbers are almost identical except for $k_2,$
which is larger for the diluted core
($k_2=0.5378$ versus $k_2=0.5287$),
given its smoother density profile.
Even with the high-quality data returned by the Juno
mission, the degree of mixing of the core is still 
under investigation \citep{Stevenson2018}.

\section{Discussion}\label{Sec__Discussion}
The possibility of improving our knowledge of the interior structure
of exoplanets beyond the information provided by the mass and radius
rests on the availability of additional data regarding atmospheric 
composition \citep[e.g.,][]{Madhusudhan2016}, 
stellar mass and composition \citep[e.g.,][]{Dorn2015}, and measurement of the
fluid Love number $k_2$ (possibly of $k_{n}$ for $n>2$). 
While some or all of these parameters are expected to be measured in 
the future for an increasing number of extrasolar planets, 
$k_2$ is the most direct constraint 
on the interior structure, given its dependence
on the density profile.

\subsection{Nonlinear effects in the calculation of the Love numbers}
\label{Sec__DiscussionNonLinearity}
The Love numbers as defined here depend only on
the density profile, and thus, they represent intrinsic properties
of a given planet, much like the case of the moment of 
inertia defined in equation (\ref{Eq__MoI}).
From the knowledge of the Love numbers, the
response of the planet to a given perturbation can be 
determined.
In the case of tidal perturbations, we would apply equations 
(\ref{Eq__TidalPotential}) and (\ref{Eq__TidalRadialDeformation})
to estimate the tidally induced modification of 
the gravitational field and of the
shape of the surface.
A similar approach applies to the case of rotational perturbation
(section \ref{Sec__SurfaceDeformation}).
Thus, under the assumption of linear response
(equation \ref{Eq__TidalPotential}) and of 
linear combination of different perturbations 
(e.g., equation (\ref{Eq__LinearCombination})), from the 
observation of the gravitational field and/or the 
shape, and by knowing the parameters of the perturbing
potential, we may be able
to invert for the Love numbers,
and thus for the degree of concentration 
of mass in the interior.
This simple strategy would not be accurate if the response
were not linear, that is, if a given degree $n$ of the perturbing
potential induced a response in a degree $q\neq n$, and
if the perturbations induced by different processes, typically,
rotation and tides, could not be added linearly.
An example of a nonlinear response is provided by the Earth, 
where the degree-2 tidal perturbations
of the Sun and the Moon induce a response in the degree-4
gravitational harmonics.
However, the amplitude is more than three orders of magnitude
smaller than the corresponding correction for $n=2$ \citep{Petit2010}.
When the rotational perturbation is much stronger than
the tidal perturbation, nonlinear effects appear, which
induce an increase of
$k_2$ by about 10\% for the fast 
rotators Jupiter and Saturn \citep{Wahl2016,Wahl2017b}.
These nonlinear effects can be accurately 
estimated with the method of the concentric 
Maclaurin spheroids \citep{Hubbard2013}, an approach that
requires more involved computations 
than the propagator-matrix
developed here.

The appearance of nonlinear effects make
the Love numbers dependent on both the planet
interior structure and the dynamical
environment of the planet.
Thus, they are no longer a fundamental 
property of the planet.
In this work we focused on the linear theory.
This approach has the main advantage 
that the Love numbers do
represent a measure of the internal concentration
of mass (Figure \ref{fig__GJ1214b}), 
they can be straightly compared among different objects
(Figures \ref{Fig__GJ436b} and \ref{Fig__Jupiter}),
and their computation is quite fast 
(Figure \ref{Fig__k2ComputationTime}).
In addition, tidally locked hot Jupiters represent
one of the best targets for the measurement
of $k_2$ \citep{Ragozzine2009,Batygin2009}, and their
rotational and tidal perturbations are comparable,
thus making the linear theory applicable \citep{Ragozzine2009,Wahl2017b}.

\subsection{Observability and interpretation}
\label{Sec__DiscussionInterpretation}
The value of $k_{n}$ can be obtained by inverting the transit light
curve for the shape of the surface of the planet
\citep{Correia2014,Hellard2018}, which can 
be modeled with the Love numbers $h_{n}$
and an expression for the radial surface deformation
(e.g., Eq. (\ref{Eq__LinearCombination})).
Under the assumption of hydrostatic equilibrium, which 
assumes that the surface of the 
planet represents an equipotential surface corresponding to the
body's tidal and rotational
potentials, there is a simple relation between the Love numbers  $h_{n}$ and 
the Love numbers $k_{n}$ \citep[Section \ref{Sec__Intro} and, e.g.,][]{Munk1960}:
\begin{equation}
h_{n}=1+k_{n}.
\end{equation}
Thus, the determination of $h_{n}$ would simply translate
into the determination of $k_{n}$, which provides the additional 
constraint on the interior structure.

The tidal and rotational potentials that modify the shape of the 
planet induce a related modification of the gravitational field, which in 
turn modifies the gravitational
interaction of the planet with the parent star and
with additional planets in the system, if any are present.
In general, this modification
results in orbital perturbations and evolution.
The evolution is associated with dissipative processes,
which depend on the Love number $k_2$
and the dissipation parameter $Q$ of
both the star and the planet \citep[e.g.,][]{Goldreich1966,Lainey2017}.
However, the variation of the orbital parameters occurs on a variety
of timescales, and for some specific dynamical configurations,
there exist fast components, like the apsidal precession,
that do not depend on the dissipation parameters and are 
controlled by the value of $k_2$
of the planet \citep{Mardling2007,Ragozzine2009,Batygin2009}.
For such cases, even with the relatively short
temporal baseline of extrasolar planet
observations at the present time, 
there are some first successful attempts 
at constraining or placing bounds on
the values of $k_2$ for some exoplanets
\citep{Batygin2009,Csizmadia2018}.

Independent of the method used to infer $k_2$,
the possibility of 
interpreting its value
in terms of the interior structure rests on the assumption that the 
planet is relaxed.
Since gases do not have shear strength, gaseous planets
should satisfy the hypothesis of hydrostatic equilibrium.
All the odd zonal harmonics of the gravitational field of the gas giant Jupiter, if in perfect hydrostatic equilibrium,
would be equal to zero.
However, its gravitational field is north-south asymmetric
\citep{Iess2018}, and this non-hydrostatic component
is informative of the wind dynamics
\citep{Kaspi2018}.
The non-hydrostatic component also modifies the 
even zonal harmonics, whose interpretation is thus
affected by the depth of the wind dynamics
\citep{Guillot2018}.
Still, hydrostatic models represent the starting point
for investigating its interior \citep{Wahl2017,Guillot2018}.
In the foreseeable future, there is no possibility 
of inferring the high-order
spherical harmonics of the gravitational fields of
extrasolar planets, thus making hydrostatic models
the starting (and likely ending) point for the modeling
of their interior structure.

Terrestrial planets are objects whose main 
constituents are metals, rocks, and ices.
These materials, unlike gases, have finite shear strengths
and do not respond instantaneously to a 
perturbing potential.
Their response is a function of the material 
properties, which in general are strongly affected by the
temperature, and of the timescale and history of the perturbation
\citep[e.g.,][]{Padovan2014}.
Thus, the assumption of hydrostatic equilibrium 
for this class of objects
requires an assessment of their orbital history
and global internal evolution.

\section{Conclusions}
We provided a semi-analytical method for computing the fluid Love
numbers $k_{n}$ of any planet from the 
knowledge of its density profile (section \ref{Sec__MatrixPropagator}).
This parameter depends on the radial distribution of mass, 
thus providing additional information on the interior structure
beyond the mass and radius.
The method has been benchmarked against several results 
available in the literature (section \ref{Sec__Validation}).
The computation is very fast (Figure \ref{Fig__k2ComputationTime}), and
the code is freely available
(\href{https://bitbucket.org/sebastianopadovan/planetary-fluid-love-numbers/src/master/}{https://bitbucket.org/sebastianopadovan/planetary-fluid-love-numbers/src/master/}).
We used a few simple cases to investigate the basic
dependencies of the Love numbers
on the interior structure of a planet (Figure \ref{fig__GJ1214b}).
In Figure \ref{Fig__GJ436b} we illustrate the application of 
the code to the planet \object{GJ 436b}, whose
observed mass and radius are compatible with an ocean-like
interior
or an interior dominated by rocks and metals 
\citep{Adams2008}, and in Figure \ref{Fig__Jupiter} 
we apply the code to a Jupiter-like hot Jupiter, whose core might be
diluted due to erosion during the age of the solar system
\citep{Wahl2017}.
These basic applications show that 
measuring
$k_2$ would improve our understanding of the interior
of extrasolar planets, but of course even a perfect
knowledge of its value would not completely
remove the
degeneracy of planetary interior models 
\citep[e.g.,][]{Kramm2011}.

In applying the calculation of the fluid Love
numbers presented here, the following points are 
worth noting:
\begin{itemize}
        \item[1. ] The fluid Love numbers obtained from solving
    Eq. (\ref{Eq__PotentialEquation}) describe the tidal
    response of a fluid nonrotating planet. The theory
    can also be used to describe the deformation of a 
    planet in a state of rotation synchronous with
    its circular orbit, that is, a planet for which the 
    rotational frequency $\omega$ in Eq. 
    (\ref{Eq__RotationalRadialDeformation}) corresponds 
    to the orbital frequency $n=\sqrt[]{GM_{\rm S}/d^3}$.
    In this case, the response corresponds to the linear
    combination of the perturbing potentials $W$ and $Z$,
    as in Eq. (\ref{Eq__LinearCombination});
    
    \item[2. ] Planets that rotate more rapidly than
    the tidal perturbation orbital frequency and in which
    the rotational distortion is much larger than the 
    tidal distortion (e.g., Jupiter) cannot be treated
    with a linear theory for two reasons. 
    First, the theory presented here can only treat 
    small perturbations to a spherical planet (as a 
    reference, there is about a $6.5\%$ variation between
    the polar and equatorial radii of Jupiter, compared to
    0.3\% in the case of the Earth). Second, a dynamical
    theory of tides may be required to fully treat the 
    response when the tidal bulge moves rapidly in the 
    planet's corotating frame, depending on the proximity 
    of the tidal frequency to the planet's resonant 
    frequencies \citep[see, e.g.,][and references 
    therein]{Wahl2016};
    
    \item[3. ] The response of planets with solid layers 
    depends in general on the timescale of the perturbation and
    on the density, rigidity, and viscosity of their interiors
    \citep[e.g.,][]{Moore2000,Padovan2014}. 
    In general, the
    theory applied here cannot be used for these objects.
    However, it is possible that for a given perturbation
    the planet responds as fluid, as in 
    the case of the rotational flattening of the Earth
    \citep{Lambeck1980}, and in such case an 
    observed $k_2$ may thus be interpreted 
    with the theory applied here.
    However, it is not known 
    a priori whether an observed response for extrasolar planets corresponds to a fluid response, and an assessment of the orbital 
    and thermal history of each object would be required
    (Section \ref{Sec__DiscussionInterpretation});
    
    \item[4. ] Given the previous points, the theory
    developed here
    is applicable to close-in, tidally locked gaseous exoplanets. 
    From an observational point of view, these objects are 
    the first objects for which estimates
    of the Love number $k_2$ will become available 
    \citep[e.g.,][]{Hellard2018}.
\end{itemize}


\begin{acknowledgements}
This work was supported by the DFG within
the Research Unit FOR 2440 ``Matter Under
Planetary Interior Conditions''.
P.B. and N.T. acknowledge the support
of the DFG priority program SPP1992 
``Exploring the diversity of extrasolar planets'' (TO 704/3-1).
Figures were created with the
software described in \citet{Hunter07}.
S.P. thanks W.B. Moore and D.J. Stevenson
for informative discussions.
The feedback of an anonymous referee improved the clarity of the 
paper.
\end{acknowledgements}

\bibliographystyle{aa} 
\bibliography{Biblio}

\listofobjects

\end{document}